\documentclass{jfm}
\usepackage{graphicx}
\usepackage{epstopdf, epsfig}
\usepackage{makeidx}
\usepackage{afterpage}
\usepackage{amsmath}
\usepackage{mathtools}
\usepackage{subfiles}
\usepackage{caption,subcaption}
\usepackage{natbib}
\usepackage{url}

\shorttitle{$CO_{2}$ exchange with wave breaking}
\shortauthor{S.Li, A.V.Babanin, F.Qiao, D.Dai, S.Jiang, C.Guan}

\title{Laboratory experiments on $CO_{2}$ gas exchange with wave breaking}

\author{Shuo Li\aff{1}
  \corresp{\email{leeshuo1991@gmail.com}},
  Alexander V. Babanin\aff{1,2},
  Fangli Qiao\aff{2,3}
  Dejun Dai\aff{3},
  Shumin Jiang\aff{3},
  Changlong Guan\aff{4}
}

\affiliation{\aff{1}Department of infrastructure Engineering, University of Melbourne,
Melbourne, Victoria, Australia
\aff{2}Laboratory for Regional Oceanography and Numerical Modeling, National Laboratory for Marine Science and Technology, Qingdao, 266237, China
\aff{3}First Institute of Oceanography, Ministry of Natural Resources, Qingdao, China
\aff{4}College of Oceanic and Atmospheric Sciences, Ocean University of
China, Qingdao, China
}

\begin{document}

\maketitle

\begin{abstract}
The $CO_2$ gas transfer velocity ($K_{CO_2}$) at air-water interface in a wind-wave flume was estimated at the circumstance of wave breaking. Three types of dynamic processes in the flume were created: monochromatic waves generated by wavemaker, mechanically-generated monochromatic waves with superimposed wind forcing, pure wind waves with 10-meter wind speed ranging from 4.5 m/s to 15.5 m/s. Without wind forcing, $K_{CO_2}$ correlated with the wave breaking probability, wave height of breakers and energy loss due to wave breaking. With superimposed wind, wind speed was found to influence $K_{CO_2}$ both in the coupled wind/mechanical wave experiments and in pure wind waves, but wave breaking still played a significant role in $CO_2$ gas exchange. Therefore, wave properties should be considered directly in parameterization of $K_{CO_2}$. A non-dimensional empirical formula was established in which $K_{CO_2}$ is expressed as a function of wave breaking probability, a modified Reynolds number and an enhancement factor to account for wind speed. 
\end{abstract}


\section{Introduction}
Atmospheric carbon dioxide has been observed to be increasing in the past few decades, which exerts impact on global climate change and carbon cycle \citep{Pachauri2014}. Ocean, however, is one of the largest reservoirs for $CO_{2}$ which is a sparingly soluble gas, and has a potential to accumulate or decrease through gas exchange across air-sea interface. The $CO_2$ flux ($F$) between the atmosphere and ocean is typically described as the product of gas transfer velocity ($K_{CO_2}$), solubility ($s$) and thermodynamic driving force in terms of partial pressure difference:
\begin{equation}
F = K_{CO_{2}}\cdot s\cdot (pCO_{2w} - pCO_{2a}),
\label{eq:fkp}
\end{equation}
where $pCO_{2w}$ and $pCO_{2a}$ denote the water-side and air-side $CO_{2}$ partial pressure respectively. The rate of gas exchange ($K_{CO_2}$) is a kinetic function of environmental forcing factors such as wind speed, wave properties (height, steepness, rate and severity of breaking) and bubble production (size and amount – both related to the breaking). Because of significance of $K_{CO_{2}}$ for the issue of $CO_{2}$ exchange, particularly in the climate context, its parameterization has been a major research topic for years. Commonly, $K_{CO_{2}}$ is linked directly to the wind speed through a linear, quadratic or cubic relation \citep{Wanninkhof2009}. However, uncertainties in the relationships imply that wind speed alone is not sufficient to quantify gas transfer velocity. Especially for sparingly soluble gas such as $CO_{2}$, the efficiency of gas transfer depends on water-side resistance which decreases with more turbulent hydrodynamic processes, and in presence of waves there are turbulence induced by wave-orbital motion and by wave breaking \citep[e.g.][]{Babanin2011}. The latter also produces bubble clouds, bubbles dramatically increase the water-air interface area and, importantly, are transported down into the water column.

Original gas transfer models with water-side turbulent dissipation were proposed by \citet{Fortescue1967} and \citet{Lamont1970}. Considering the water surface waves, \citet{Jahne1987} found that the mean square slope of the waves was an appropriate parameter to describe the gas transfer velocity. In addition, the dependence of gas transfer rate on Schmidt Number ($Sc$), which is the ratio of fluid kinematic viscosity and mass diffusivity, changed from $Sc^{-\frac{2}{3}}$ to $Sc^{-\frac{1}{2}}$ at wavy surface. \citet{Zappa2004} employed infrared technique to detect microwave breaking in wave tank and found that the gas transfer velocity scaled well with fractional area coverage of microbreakers. \citet{Zhao2003} attempted to relate gas transfer velocity to the sea whitecap coverage by using a wind-sea Reynolds number which represented the turbulence generated by waves. The relationship between gas transfer and wind-sea Reynolds numbers was further evaluated in \citet{Brumer2017} with field data. The bubble effect on gas exchange was also recognized in previous studies \citep{Woolf1997,Liang2013}. Bubble-mediated gas transfer was generally parameterized with wind speed, but it is obvious that the bubbles (except in hurricane-like condition) are produced by wave breaking rather than by the wind. The COARE model in \citet{Fairall2011} should also be mentioned, which combines various mechanisms of the gas exchange.

Thus, the wave dissipation process due to breaking is a most essential subject of surface wave dynamics, relevant for the gas exchange. The importance of wave breaking on air-sea interaction has been discussed in \citet{Melville1996} and \citet{Babanin2011}. Due to the lost energy, breaking enhances intensity of the under-surface turbulence by up to 3 orders of magnitude, it produces bubbles and may spend up to 50\% of energy loss on work against the buoyancy forces acting on these bubbles. Wave growth and ultimately its breaking are connected with the wind, hence there is correlation between $K_{CO_{2}}$ and wind speed, but this is by far not a direct connection and reason for the breaking is nonlinear evolution of waves (or wave superposition), not the wind \citep{Babanin2011}. \citet{Banner2000} and \citet{Babanin2001} studied the dominant wave breaking statistics in deep and finite depth water. The dominant wave breaking probability was found to be a function of significant wave steepness. \citet{Babanin2010} conducted numerical and experimental research on the breaking onset of two-dimensional steep waves. The features of wave breaking were discussed for cases with and without superimposed wind.

Since wave breaking largely facilitates gas flux across air-sea interface, we conducted laboratory experiments to investigate how $CO_{2}$ gas exchange vary with wave breaking. Experimental setup is introduced in section~\ref{sec:the_exp}. Section~\ref{sec:results} is dedicated to the analysis of relationship between $CO_{2}$ exchange velocity and environmental forcing factors. Further discussions and conclusions are presented in section~\ref{sec:discussion}.

\section{The Experiments}\label{sec:the_exp}
The facility for experiments was a wind-wave flume, 45 m long, 1.8 m high and 1 m wide available at First Institute of Oceanography in China. The tank was filled with tap water up to 1.2 m. The wind fan is installed above the wave tank with closed air channel. A mechanical wavemaker is located upstream. It is programmable and able to generate regular waves, steep enough to lead to wave breaking. At the downstream end of wave tank, a beach was used for damping wave energy (more than 95\%) to prevent the reflection of waves.

\begin{figure}
	\centering
	\includegraphics[width=10cm]{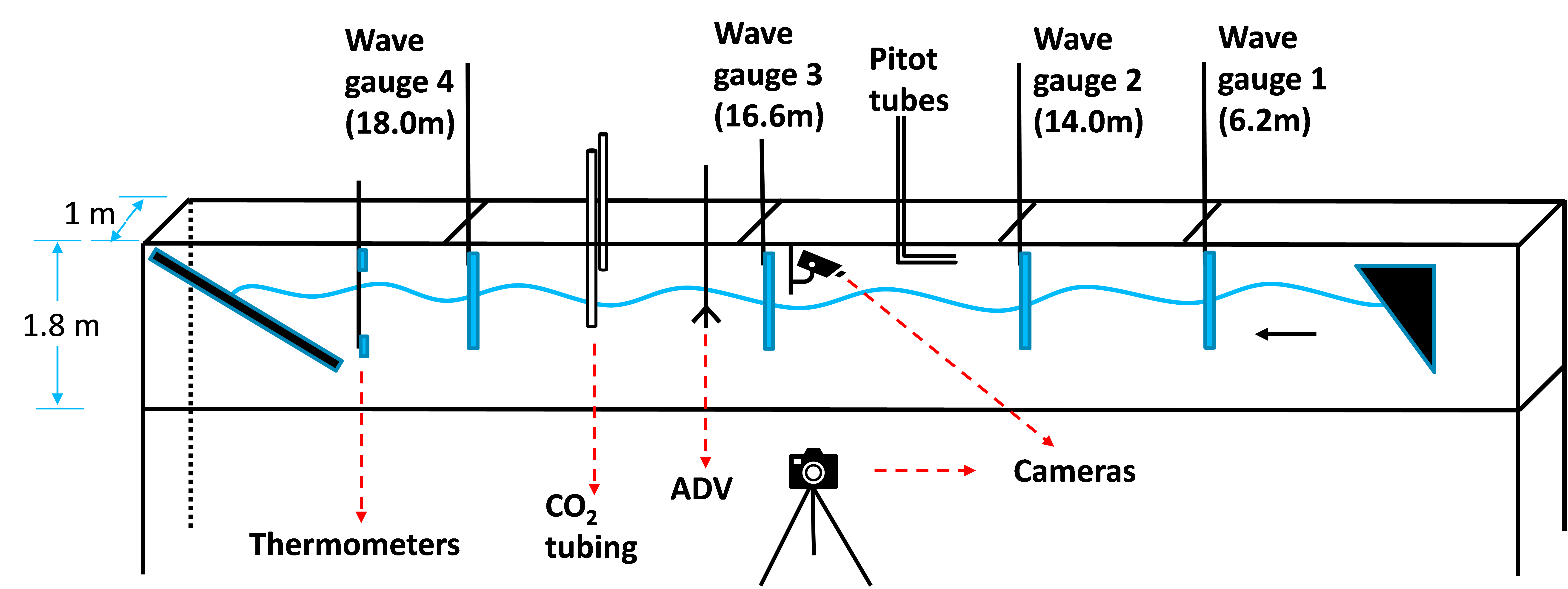}
	\caption{Schematic of deployment of probes in the wave tank. There are four resistance-type wave gauges at 6.2m, 14.0m, 16.6m, 18.0m from the wavemaker on the right. Close to wave gauge 3, a set of Pitot tubes, one ADV and sampling tubing for $CO_2$ analysis were installed. Outside the wave tank, a Canon camera and a video camera were used to record waves. At the downstream rear of wave tank, a pole with two thermometers attached was placed, for air and water temperature respectively.}
	\label{fig:setup2}
\end{figure}
Various sensors were employed along the wave tank to measure physical and chemical properties. Water surface elevations were measured by 4 resistance-type wave gauges (figure~\ref{fig:setup2}) at 50 $Hz$ sampling rate located at 6.2 m, 14.0 m, 16.6 m and 18.0 m from the wavemaker. A vertical array of 5 Pitot tubes was located about 10 cm before wave gauge 3, arranged evenly (5 cm spaced) with the lowest one at about 15 cm above the free water surface. It took 100 milliseconds for the computer to record wind speed at each tube. An acoustic Doppler velocimeter (ADV) was installed side-by-side with wave gauge 3 to measure turbulence in the water, although this data is not used in the present work. 50 cm downstream of wave gauge 3, tubing for taking water and air samples in the flume was installed, and further connected to the $CO_{2}$ analysis devices. Two thermometers were placed at the rear of wave tank for air and water temperature measurements respectively. Air conditioners in the lab were always running during experiments so that temperature at different locations of wave tank was almost the same. Outside of the wave tank, a Canon digital camera and a video camera were employed to record wave breaking processes. In addition, the water acidity index ($pH$) and air pressure in the lab were also recorded during the experiments.

The instrument for $CO_{2}$ analysis was Apollo (AS-P2 by Apollo SciTech, USA) which was incorporated with air-water equilibrator and Picarro G2301 analyzer. Water was piped out of tank at a rate of around 2.5 L/min into equilibrator to contact with air stream. After reaching equilibrium, the water was returned to the rear of tank and the equilibrated gas was analyzed by Picarro as $pCO_{2w}$ in equation~(\ref{eq:fkp}). The multi-position valve on Apollo was set up so that Picarro could analyze equilibrated gas samples and ambient air in lab alternatively. Meanwhile, Apollo was able to collect data of both $pCO_{2w}$ and $pCO_{2a}$ from Picarro. A drying section was assembled for Picarro to absorb water vapor in gas stream. Standard gases with $CO_{2}$ concentration of 400.0 ppm, 600.7 ppm, 799.2 ppm, and 1000.6 ppm were used to calibrate Picarro since there might be drift of measurements. In the flume, a $pCO_{2}$ disparity was created between air and water at the beginning of each experiment by adjusting $CO_{2}$ concentration in water to be bigger than that in air. So $CO_{2}$ would escape from water during the occurrence of wave breaking.

\begin{table}
	\scriptsize
	\begin{center}
		\begin{tabular}{ccccccccccccccc}
			
			Case&$f_{0}$&$a_{0}$&$\varepsilon_{0}$&$f_{+}$&$a_{+}$&$BFI$&$f_{fan}$&$U_{10}$&$H_b$&$U_{wb}$&$b_{T}$&$H_s$&$U_{wm}$&$K_{600}$\\
			$No.$&$(Hz)$&$(m)$& &$(Hz)$&$(m)$& &$(Hz)$&$(m/s)$&$(m)$&$(m/s)$& &$(m)$&$(m/s)$&$( 10^{-6} m/s)$\\
			
			A1&1.2 & 0.035 & 0.20 & 1.32 & 0.010 & 0.95 & 0 &0&0.15&0.60&0.092&0.32&0.13&1.065\\
			
			A2&1.2 & 0.052 & 0.30 & 1.33 & 0.006 & 1.36 & 0 &0&0.18&0.70&0.103&0.48&0.19&1.386\\
			
			A3&1.0 & 0.050 & 0.20 & 1.10 & 0.024 & 0.95 & 0 &0&0.25&0.82&0.073&0.40&0.19&1.280\\
			
			A4&1.3 & 0.029 & 0.20 & 1.43 & 0.007 & 0.95 & 0&0&0.10&0.36&0.101&0.30&0.12&0.748\\
			
			A5&1.1 & 0.041 & 0.20 & 1.21 & 0.014 & 0.95 & 0&0&0.23&0.81&0.090&0.35&0.16&1.492\\
			
			A6&0.9 & 0.061 & 0.20 & 1.04 & 0.035 & 0.60 & 0&0&0.30&0.90&0.069&0.43&0.22&1.713\\
			
			A7&1.1 & 0.033 & 0.16 & 1.24 & 0.019 & 0.59 & 0&0&0.15&0.53&0.111&0.31&0.14&1.206\\
			
			A8&1.1 & 0.051 & 0.25 & 1.24 & 0.014 & 0.94 & 0&0&0.20&0.72&0.111&0.42&0.19&1.396\\
			
			A9&1.0 & 0.055 & 0.22 & 1.11 & 0.023 & 0.95 & 0&0&0.24&0.76&0.098&0.42&0.20&1.663\\
			
			A10&0.9 & 0.055 & 0.18 & 1.02 & 0.039 & 0.61 & 0&0&0.29&0.87&0.120&0.40&0.21&2.946\\
			
			B1&0.9 & 0.055& 0.18 & 1.02 & 0.039 & --- & 25&11.21&0.29&0.85&0.121&0.46&0.24&4.790\\
			
			B2&0.9 & 0.055& 0.18 & 1.02 & 0.039 & --- & 15&6.77&0.28&0.82&0.122&0.42&0.22&1.946\\
			
			B3&1.1 & 0.041& 0.20 & 1.21 & 0.014 & --- & 20&9.14&0.20&0.74&0.087&0.40&0.18&2.997\\
			
			B4&1.1 & 0.041& 0.20 & 1.21 & 0.014 & --- & 30&13.43&0.24&0.88&0.086&0.52&0.22&4.101\\
			
			B5&1.0 & 0.055& 0.22 & 1.11 & 0.023 & --- & 20&8.85&0.26&0.86&0.098&0.46&0.22&3.898\\
			
			B6&1.0 & 0.055& 0.22 & 1.11 & 0.023 & --- & 30&13.43&0.27&0.88&0.096&0.55&0.25&6.974\\
			
			C1& --- & --- & --- & --- & --- & --- & 10 &4.46&0.02&0.21&0.239&0.15&0.02&0.092\\
			
			C2& --- & --- & --- & --- & --- & --- & 15 &6.88&0.03&0.29&0.371&0.22&0.04&0.280\\
			
			C3& --- & --- & --- & --- & --- & --- & 20 &9.19&0.04&0.34&0.488&0.27&0.05&0.670\\
			
			C4& --- & --- & --- & --- & --- & --- & 25 &11.12&0.05&0.38&0.559&0.31&0.07&0.999\\
			
			C5& --- & --- & --- & --- & --- & --- & 30 &13.25&0.07&0.44&0.665&0.39&0.09&---\\
			
			C6& --- & --- & --- & --- & --- & --- & 35 &15.44&0.09&0.51&0.732&0.45&0.12&2.743\\
			
		\end{tabular}
		\caption{Experimental parameters of all tests. A1 to A10 are the monochromatic experiments with mechanically-generated waves. B1 to B6 are the coupled wave experiments (wind forcing the mechanically-generated waves). C1 to C6 represent the wind-wave experiments. }
		\label{tbl:expinput}
	\end{center}
\end{table}
The experimental parameters are listed in table~\ref{tbl:expinput}. Three kind of experiments were conducted in terms of wave generation method: monochromatic waves generated by mechanical wavemaker (A1-A10), mechanically-generated waves coupled with superimposed wind (B1-B6), waves produced and forced by wind only (C1-C6). The initial wave signal of mechanical waves in A1 to B6 was the combination of a carrier sinusoidal wave with frequency $f_0$, amplitude $a_0$, wave number $k_0=(2\pi f_0)^2/g$,where $g$ is gravitational acceleration, steepness $\varepsilon_0=a_0 k_0$, and a resonant sideband with frequency $f_+$, amplitude $a_+$ (10\% to 30\% with respect to $a_0$). The Benjamin-Feir Index ($BFI$) was used to evaluate the instability of wave trains, as $BFI=\varepsilon_0 /(\Delta k/k_0)$, where $\Delta k$ is the wave number difference between carrier wave and sideband. The frequency of wind fan ($f_{fan}$) in B1-C6 was set up beween 10 $Hz$ to 35 $Hz$. 10-meter wind speed $U_{10}$ was computed by using drag coefficient $c_d=0.0013$ in the lab and wind friction velocity calculated from the measurements of Pitot tubes. For cases A1 to B6, propagation of the modulated mechanical waves was unstable and led to breaking after passing wave gauge 2. The records of wave gauge 2 to 4 were used to recognize and quantify the breakers (e.g. the energy loss) because an evident decrease of the wave height after the breaking was observed. In addition, the identified breakers were also confirmed with videos. By choosing the breaking events that happened upstream of the nearest to $CO_2$ sampling tubing, the wave height measured at wave gauge 2 or 3, before the breaking, was used as the proxy of wave height of the breaking onset, $H_b$ (in the table this is mean value of the measured breaking wave height over the repeated runs). Similarly, $U_{wb}$ was the mean breaking wave orbital velocity which is the product of wave amplitude and angular frequency, following the linear wave theory. For relatively short wind waves (C1-C6), the breaking events were identified by using the criterion for ultimate steepness of individual waves in the monochromatic waves subject to modulational instability $\varepsilon = 0.44$ \citep{Babanin2007,Babanin2010}. $b_T$ of every case was estimated as the proportion of the number of recognized breakers among the count of all waves. Significant wave height $H_s$ and mean orbital velocity $U_{wm}$ of all waves at wave gauge 3 were also estimated. Plunging or spilling breakers were found for mechanically-generated waves, while microbreaking dominated the young and short wind-produced waves due to the limited fetch.

The method to calculate gas transfer rate $K_{CO_2}$, following \citet{Donelan1994}, was adopted in our work:
\begin{equation}
\frac{\partial C_{g}}{\partial t}\frac{V_{w}}{A}=-K_{CO_{2}}(C_{g}-C_{a}),
\label{eq:md_lab_2}
\end{equation}
where $C_{g} $ and $C_{a}$ are $CO_{2}$ concentration in equilibrated gas and in air, respectively. Here,  $V_{w}$ and $A$ is the water volume and surface area that are involved with the gas exchange processes. So, $V_{w}/A$ identifies the height of water column, related to the depth of turbulent mixing layer. \citet{Thomson2016} suggested that the turbulence could be transported down to wave trough due to orbital motion, and in our work, the depth of upper mixed layer was scaled with $H_b$. The calculated $K_{CO_2}$ was further corrected to 20$^{o}C$ of fresh water with Schmidt number $Sc_{600}=600$ in order to eliminate the thermal effect on gas transfer.
\begin{equation}
\frac{K_{CO_2}}{K_{600}}=\left( \frac{Sc_{co_2}}{Sc_{600}} \right)^{-0.5},
\label{eq:k_sc}
\end{equation}
where $K_{600}$ represents the corrected transfer velocity, $Sc_{co_2}$ is the Schmidt number of water in laboratory. The magnitude of power of $Sc$ is $-0.5$ for wavy surface in the tank \citep{Jahne1987}.

\section{A new parameterization for $CO_2$ gas transfer velocity}\label{sec:results}
The estimated gas transfer velocities $K_{600}$ are listed in table~\ref{tbl:expinput}. Compared with groups of monochromatic wave experiments (A5, A9, A10), $H_s$ and $K_{600}$ in B1-B6 become bigger due to superimposed wind while $b_T$ tends to reduce with growth of wind speed. Wind forcing can slow down the modulation of unstable waves and decrease the number of breakers \citep{Babanin2010,Galchenko2012}. For wind waves, the change from microbreakers without bubble injection to large breakers with bubbles was observed when wind speed was varied from low to high.

\begin{figure}
	\centering
	\includegraphics[width=10cm]{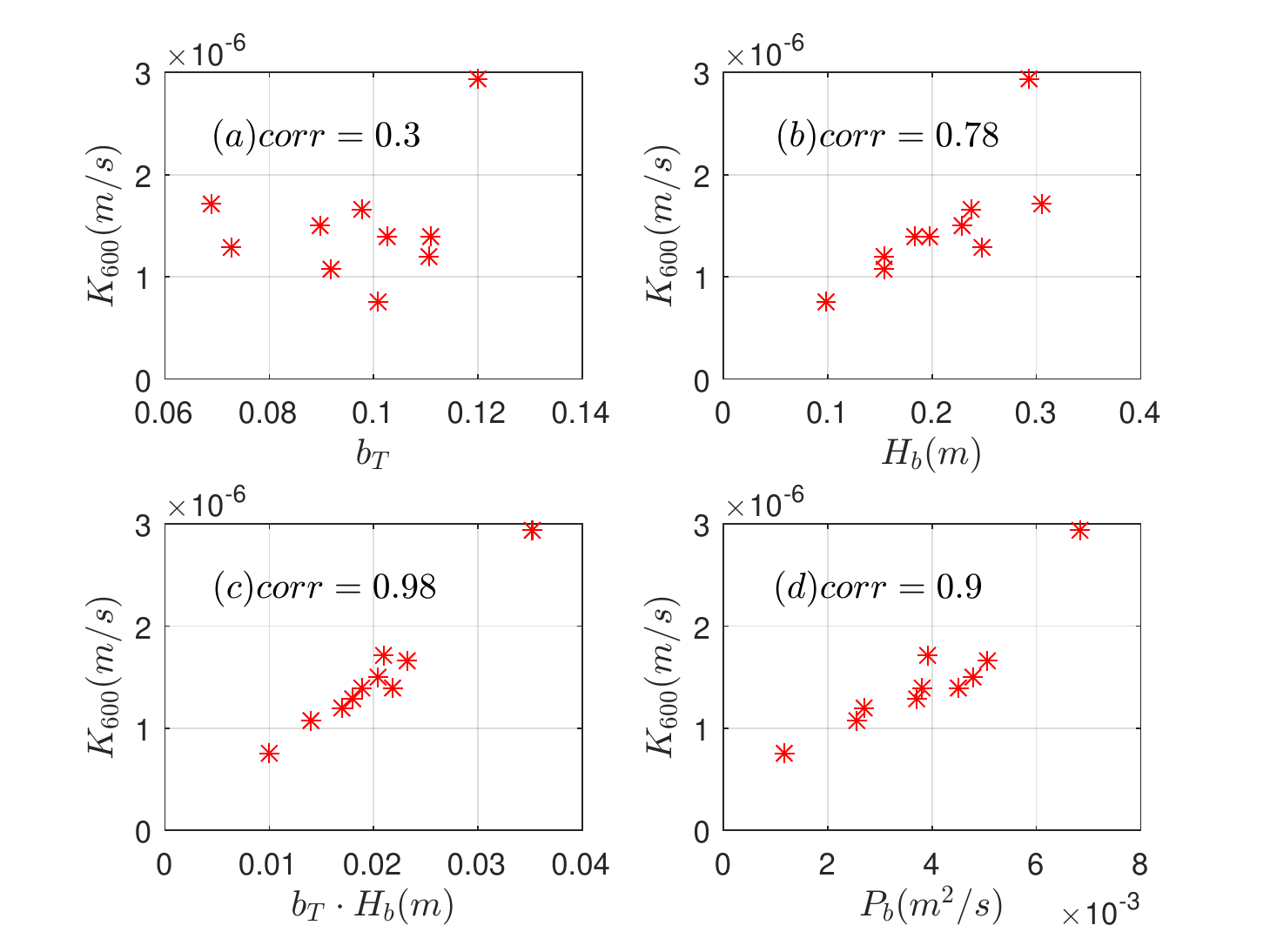}
	\caption{$CO_{2}$ gas transfer velocity of monochromatic wave experiments in correlation with (a) wave breaking probability $b_T$, (b) mean wave height of breakers $H_b$, (c) product of breaking probability and mean wave height of breakers, (d) mean energy loss of breakers upstream the nearest sampling tubing per unit of time.}
	\label{fig:mono_k_bt}
\end{figure}

In the monochromatic wave experiments (A1-A10), $K_{600}$ dependence on $b_T$ is weak (figure~\ref{fig:mono_k_bt}$(a)$), correlation between $K_{600}$ and $H_b$ is much better (figure~\ref{fig:mono_k_bt}$(b)$), but correlation of $K_{600}$ and their product $b_T\cdot H_b$ in figure~\ref{fig:mono_k_bt}$(c)$ is 98\%. The results meet our expectation because $b_T$ determines the frequency of occurrence of the water mixing events by breakers, while higher $H_b$ leads to greater orbital motion which produces more turbulence. In figure~\ref{fig:mono_k_bt}$(d)$, $K_{600}$ is also well correlated with the rate of the mean energy loss ($P_b$) within experimental periods defined by
\begin{equation}
P_b=\frac{\sum{(H_{b1}^{2}-H_{b2}^{2})}}{\Delta t},
\label{eq:deb}
\end{equation}
where $H_{b1}$ and $H_{b2}$ are the wave height before and after wave breaking measured by wave gauges, $\Delta t$ is the time length of each experiment. $P_b$ contains the information of wave breaking probability and average breaking strength, and its high correlation with $K_{600}$ is not surprising: energy lost due to breaking is then passed to the turbulence whose production rate is described by $b_T\cdot H_b$ in figure~\ref{fig:mono_k_bt}$(c)$. The results of experiments A1 to A10 demonstrate that wave breaking can still enhance $CO_{2}$ gas flux without wind, and wave characteristics are directly relevant to the $CO_{2}$ gas exchange rate. 

\begin{figure}
	\centering
	\includegraphics[width=14cm]{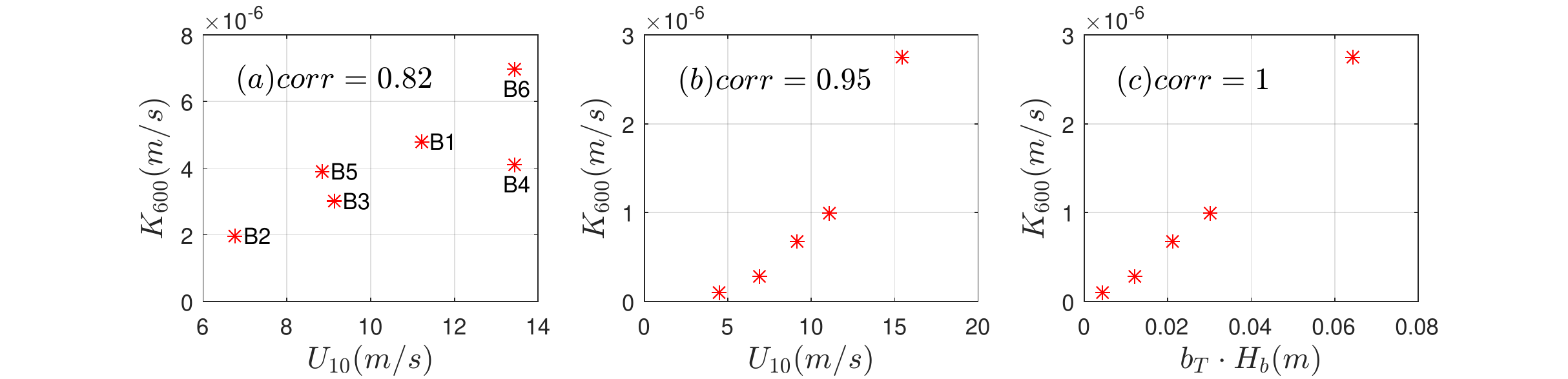}
	\caption{$CO_{2}$ transfer velocity versus 10-meter wind speed in (a) coupled wave experiments (B1-B6), (b) wind wave experiments (C1-C6). (c) $CO_{2}$ transfer velocity versus product of breaking probability and mean wave height of breakers in C1 to C6.}
	\label{fig:k_U10}
\end{figure}
Figure~\ref{fig:k_U10} shows the good correlation of $CO_{2}$ transfer velocity with 10-meter wind speed for coupled wave experiments (B1-B6) in panel$(a)$ and wind wave experiments (C1-C6) in panel$(b)$. The wind speed is a good parameter in expressing the gas transfer. However, experiments with similar $U_{10}$ can lead to different $K_{600}$ (e.g. B4 and B6 in panel$(a)$) which fact explains the uncertainties in the existing parameterization with wind speed alone. The difference of wave breaking probability in B4 and B6 (0.086 and 0.096, respectively) indicates that wave properties should also be considered. For experiments C1 to C6, $K_{600}$ is highly related with not only wind speed (panel$(b)$), but also with wave parameters $b_T\cdot H_b$ in panel$(c)$. In physical aspect, $CO_{2}$ exchange velocity is determined by water-side turbulence which in our work is related with breaking rates and turbulence originate from each breaking event. The indirect contribution of wind forcing lies in the energy input into waves and adjusting wave breaking behavior.   

\begin{figure}
	\centering
	\includegraphics[width=12cm]{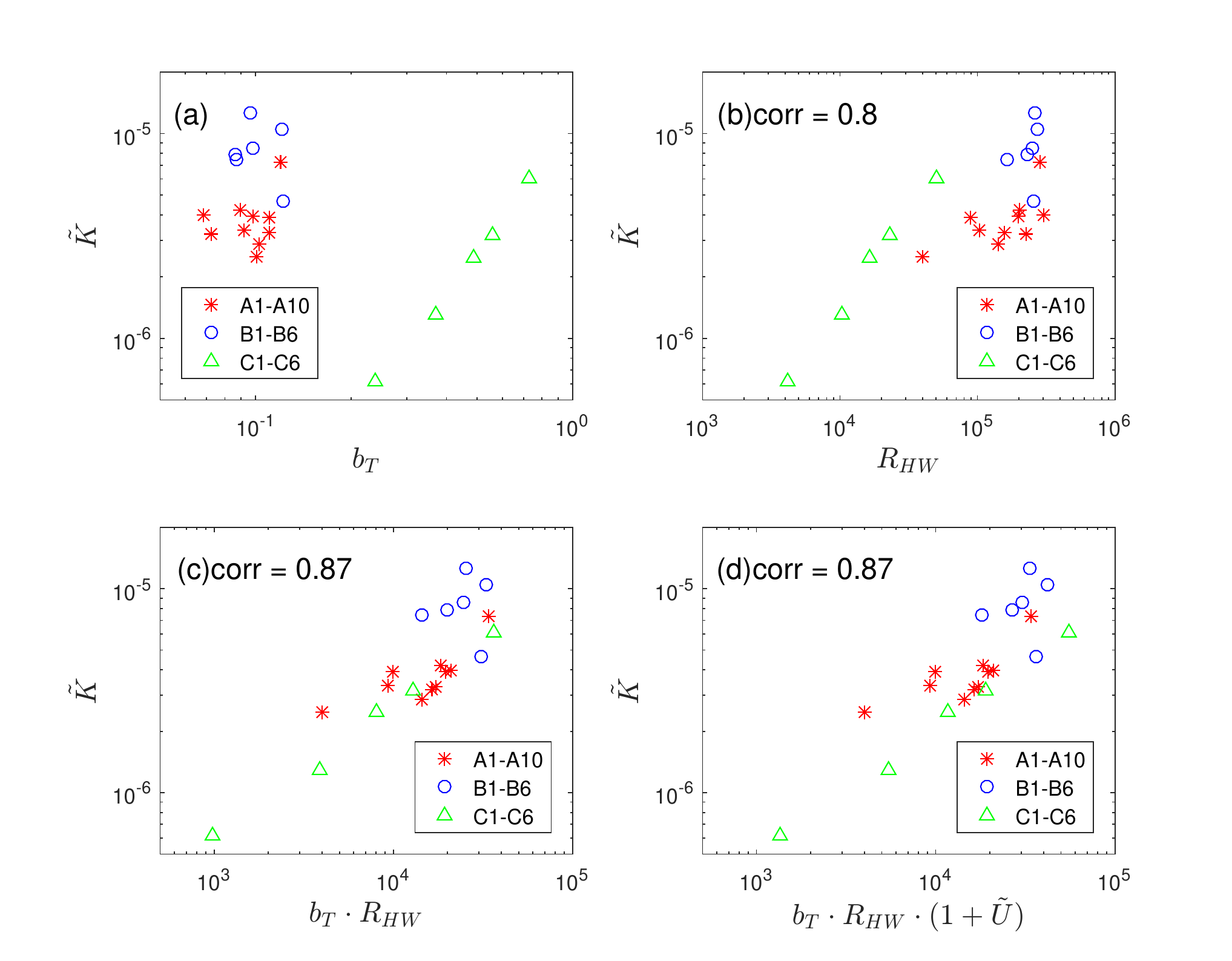}
	\caption{Non-dimensional $CO_2$ gas transfer velocity versus (a) wave breaking probability, (b) Reynolds number (\ref{eq:rhw_h}), (c) product of breaking probability and Reynolds number, (d) product of breaking probability, Reynolds number and scaled wind speed.}
	\label{fig:wind_kbtall}
\end{figure}
The parameterization of $CO_{2}$ exchange velocity should be able to unify all data sets and physically reasonable. First, considering that $K_{600}$ may be of a different order for open ocean (i.e. to avoid the dependence on the dimensional wave parameters), it is scaled by the mean orbital velocity ($U_{wm}$) of waves through (\ref{eq:nond_k_u}).
\begin{equation}
\tilde{K} = \frac{K_{600}}{U_{wm}},
\label{eq:nond_k_u}
\end{equation}
where $\tilde{K}$ is now a non-dimensional gas transfer velocity. Thus, waves are directly related with gas exchange. Meanwhile, wave height $H_{b}$ is parameterized as the form of Reynolds number which is denoted as $R_{HW}$ in (\ref{eq:rhw_h}).
\begin{equation}
R_{HW} = \frac{H_b\cdot U_{wb}}{\nu},
\label{eq:rhw_h}
\end{equation}
where $U_{wb}$ is the mean orbital velocity of breakers, $\nu$ is the viscosity of water. The transformed Reynolds number is physically relevant to wave induced turbulence. Wind speed is also scaled as non-dimensional in equation~(\ref{eq:nond_u_uf}) as was introduced in \citet{Lenain2017}.
\begin{equation}
\tilde{U} = \frac{U_*}{\sqrt{g\cdot H_s}},
\label{eq:nond_u_uf}
\end{equation}
where $\tilde{U}$ is non-dimensional wind speed, $U_*$ is the wind friction velocity, $g$ is gravitational acceleration, $H_s$ is significant wave height. In figure~\ref{fig:wind_kbtall}$(a)$, $b_T$ alone obviously can not unify the results from three data sets. In figure~\ref{fig:wind_kbtall}$(b)$, the correlation by using $R_{HW}$ is better, but the disparity between data sets is still evident. In figure~\ref{fig:wind_kbtall}$(c)$, the product of $b_T$ and $R_{HW}$ is used which signifies the importance of both wave breaking probability and wave-related turbulence. Although the correlation is improved, the data set of B1-B6 (circles) is not consistent with data set of A1-A10 (stars). The gas transfer velocity for B1-B6 is enhanced compared with that of A1-A10 due to superimposed wind while the $b_T\cdot R_{HW}$ for two data sets are similar. Thus, considering the indirect role of wind impact on $CO_2$ exchange, the scaled wind speed is applied as an enhancement factor as $(1 + \tilde{U})$ which is shown in figure~\ref{fig:wind_kbtall}$(d)$. This way, when wind forcing approaches zero, the results have to converge to the no-wind (mechanically generated) conditions. Although the correlation is the same, result in panel$(d)$ is physically more reasonable. It should also be mentioned that $\tilde{U}$ alone is unable to unify experiments B1-B6 and C1-C6 (not shown here).

The whole expression is then as following:
\begin{equation}
\tilde{K} = \alpha \cdot (b_{T}\cdot R_{HW} \cdot (1 + \tilde{U}))^{\beta},
\label{eq:nond_k_all}
\end{equation}
where $\alpha$ and $\beta$ are fitting parameters.
\begin{figure}
	\centering
	\includegraphics[width=7cm]{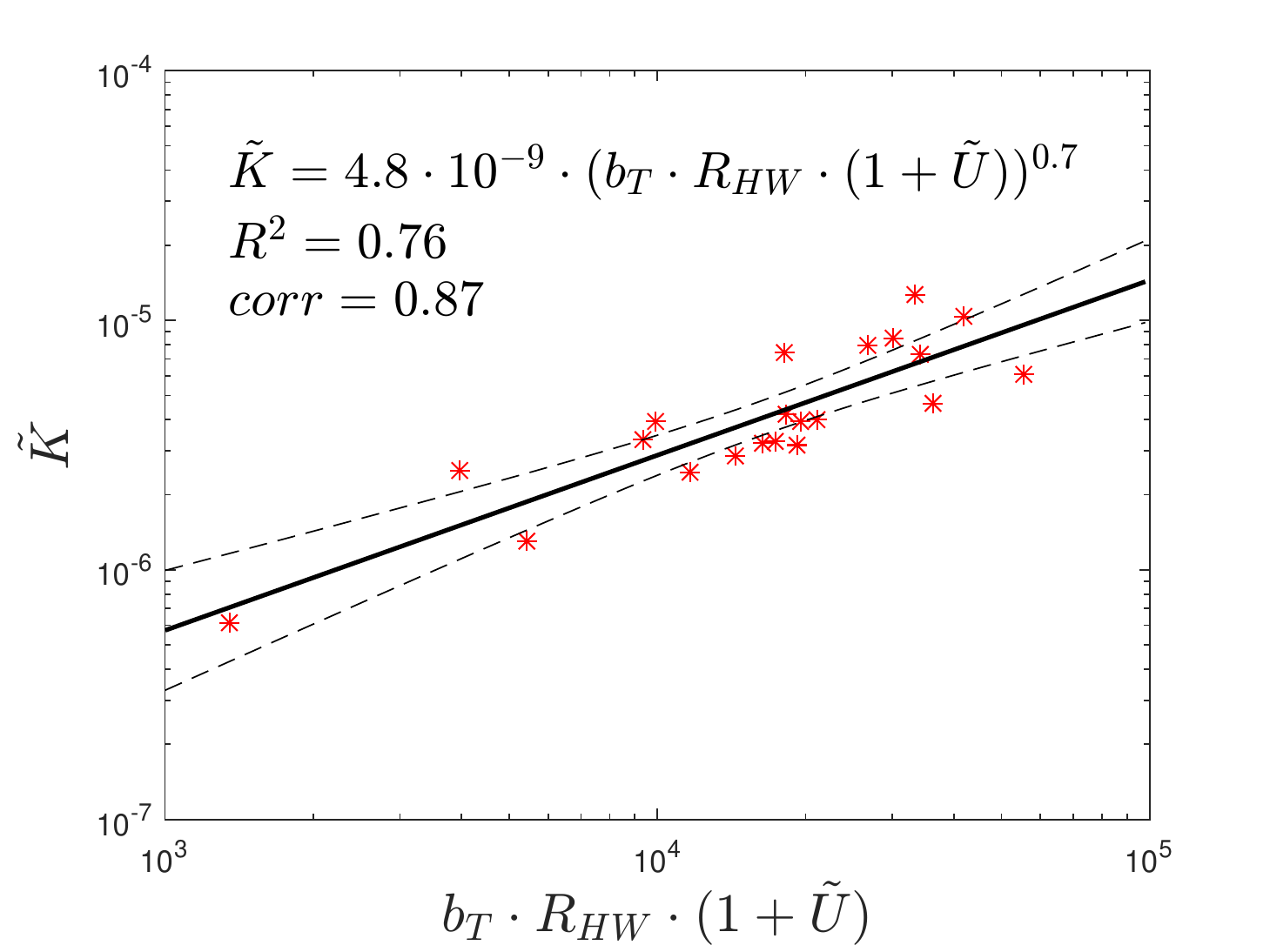}
	\caption{Logarithm power law fit for non-dimensional $CO_{2}$ gas transfer velocity of all experiments in relation with combined parameters. The dashed lines denote 95\% confidence intervals.}
	\label{fig:lab_k_all}
\end{figure}
Figure~\ref{fig:lab_k_all} shows the power-law fit between $\tilde{K}$ and combined variables. The correlation is 87\% with the coefficient of determination of 76\%. The parameters $\alpha$ and $\beta$ are $4.8\cdot 10^{-9}$ and $0.7$ respectively. The dashed lines represent 95\% confidence intervals. As mentioned above, the enhancement factor $(1 + \tilde{U})$ reduces to 1 when the wind forcing is absent so wave parameters $b_{T}$ and $R_{HW}$ are considered to have major relationship with the gas exchange. 

\section{Discussion and Conclusions}
\label{sec:discussion}
Considering the fact that water-side dynamic processes have direct impact on sparingly soluble gas (e.g. $CO_2$) exchange, we scale $k_{600}$ with the mean wave orbital velocity rather than wind speed in equation~(\ref{eq:nond_k_u}). The orbital velocity is chosen because water mass moves along with orbital motion. The parameterization of $R_{HW}$ in equation~(\ref{eq:rhw_h}) makes the use of mean wave height and orbital velocity of wave breakers. By substituting $H_b$ and $U_{wb}$ with $H_s$ and $U_{wm}$ which are easier to be obtained at field operations, the resulted final fitting parameters $\alpha$ and $\beta$ in equation~(\ref{eq:nond_k_all}) are estimated to be $1.7\cdot 10^{-8}$ and $0.61$, respectively. Equation~(\ref{eq:nond_k_all}) is established because wave breaking is believed to be the dominant factor for $CO_2$ gas exchange at sea - it is responsible both for production of bubbles and excessive amount of turbulence \citep[e.g.][]{Agrawal1992}. The formula~(\ref{eq:nond_k_all}) needs also validated by using field data.

The breaking event is often accompanied by whitecapping, with injection of bubbles which is not parameterized directly in our formula. 
No consensus has been reached on bubble size distribution and behavior injected by waves. The transformed Reynolds Number $R_{HW}$ in our work is used to denote turbulence effect which could possibly be related to bubbles, although further evidence is needed. The calculated wave energy change (breaking severity) is another parameter that can be correlated with bubbles \citep{Manasseh2006}. However, for the convenience of utilizing future field data (where breaking severity is usually not known), wave energy change due to breaking is not considered in equation~(\ref{eq:nond_k_all}).

Finally, we summarize the main findings in present work. $CO_{2}$ concentration in water decreases gradually with monochromatic wave breaking. The breaking probability, wave height, energy loss of breakers and wind speed are found to be well correlated with the gas exchange velocity. To parameterize the dependence, mean wave orbital velocity is used to scale $CO_{2}$ transfer rate. This is another indication of the direct role of waves on $CO_{2}$ gas exchange. Breaking probability and the transformed Reynolds Number are used to represent wave characteristics. The non-dimensional wind speed is employed as an enhancement factor. The proposed empirical formula fits well for data sets in our experiments. The result provides us an opportunity to evaluate $CO_{2}$ exchange through environmental wave and wind information.

\section*{Acknowledgments}
S.L. and A.V.B. were supported by the DISI Australia-China Centre through Grant ACSRF48199. A.V.B. acknowledges support from the U.S. Office of Naval Research Grant N00014-17-1-3021. F.Qiao was jointly supported by the National Natural Science Foundation of China under grants 41821004 and the International cooperation project of the China-Australia Research Centre for Maritime Engineering of Ministry of Science and Technology, China under grant 2016YFE0101400. The authors thank Dr. Ming Xin and Mr. Chao Li for valuable assistance with laboratory experiments.

\bibliographystyle{jfm}
\bibliography{labjfmref}

\end{document}